\documentclass[reviewcopy]{elsarticle}

\usepackage[reviewcopy]{adndt}
\usepackage{longtable}

\usepackage{mathptmx}

\usepackage{amsmath}

\usepackage{amssymb}

\biboptions{square,sort&compress}
\bibpunct[]{[}{]}{,}{n}{}{;}
\begin{document}

\begin{frontmatter}

\journal{Atomic Data and Nuclear Data Tables}


\title{Theoretical energy level spectra and transition data for 4p$^6$4d, 
4p$^6$4f and 4p$^5$4d$^2$ configurations of W$^{37+}$ ion}

  \author{P. Bogdanovich\corref{cor1}}
 \ead{Pavelas.Bogdanovicius@tfai.vu.lt}

  \author{R. Kisielius}

  \cortext[cor1]{Corresponding author.}

   \address{Institute of Theoretical Physics and Astronomy, Vilnius University,
 A. Go{\v s}tauto 12, LT-01108 Vilnius, Lithuania}


\date{19/05/2011} 

\begin{abstract}  
The ab initio quasirelativistic Hartree-Fock method developed 
specifically for the calculation of spectral parameters of heavy atoms and 
highly charged ions was applied to determine atomic data for tungsten ions. The 
correlation effects were included by adopting configuration interaction method. 
The Breit-Pauli approximation for quasirelativistic Hartree-Fock radial orbitals
was employed to take into account relativistic effects.  
The energy level spectra, radiative lifetimes, Lande factors $g$ were calculated 
for the $\mathrm{4p^64d}$, $\mathrm{4p^64f}$ and  $\mathrm{4p^54d^2}$ 
configurations of W$^{37+}$ ion. The atomic data, namely, the transition 
wavelengths, spontaneous emission rates and oscillator strengths for the 
electric dipole, electric quadrupole and magnetic dipole transitions among and 
within the levels of these configurations are tabulated. 

\end{abstract}

\end{frontmatter}

\newpage

\tableofcontents
\listofDtables
\listofDfigures

\section{Introduction}

Due to its unique features, the metallic tungsten is widely employed in devices 
which operate at extremely high temperatures. These properties facilitate 
adoption of tungsten in fusion reactors including internationally developed ITER
tokamak \cite{rea09,ski09}. High temperatures within fusion plasma cause 
tungsten to evaporate. Furthermore, the tungsten atoms in plasma are ionized to 
very high degrees. This process has a negative consequences on plasma since it 
makes plasma temperatures to drop significantly. 

To perform plasma processes modeling and diagnostics one needs huge amount of 
spectroscopic data describing various ionization stages of tungsten and other
atoms \cite{sum11,bie09}. Atomic spectroscopy can help to determine important
properties of plasma \cite{hut10,cle10,utt02}. Nevertheless, the spectra of the 
tungsten ions present in fusion plasma devices are still not investigated 
properly. This conclusion is evident from review \cite{kra09} where all 
available experimental wavelengths and energy level spectra for ions  from W~III
to W~LXXIV are provided.

It is evident from the analysis of \cite{kra09} and references within, that only
a few data for the spectra of tungsten ions with open 4d- shells are available. 
In present work we investigate a relatively simple system, namely W$^{37+}$. 
This ion has only one valence 4d- electron in its ground state. It follows from
\cite{kra09} that only 11 excited levels from the possible 48 ones for the 
configurations under consideration have been determined experimentally. 
On the other hand, a theoretical study of energy level spectra was presented in 
\cite{fou98}. However, this work suffers from the fact that the data
are provided just for some fifty percent of existing levels representing the
configurations under consideration. Furthermore, the transition data were 
determined for the highly excited levels only.

In present work we employ the quasirelativistic method \cite{bog06,bog07,bog08} 
developed specifically for the {\it ab initio} calculation of spectral 
parameters of heavy atoms and highly charged ions. A preliminary analysis of the
suitability of this method for the tungsten ions with open 4d- shell was 
presented in \cite{bog10}. It was demonstrated that our quasirelativistic 
approach is appropriate for the inclusion of the relativistic effects, and the 
further improvement of accuracy of the calculated data can be achieved with 
inclusion of the correlation effects.

In Section~\ref{calc} we provide a short description of our calculation method. 
The more detailed description of the applied approach and the analysis of the
convergence within the configuration interaction (CI) approximation applied for 
the inclusion of correlation effects is presented in \cite{bog11}. Obtained 
results are discussed in Section~\ref{result}.

\section{Calculation method}{\label{calc}}

\subsection{Quasirelativistic approximation}{\label{qrhf}}

We use a quasirelativistic approximation for {\it ab initio} calculations of ion
energy spectra. This approach significantly differs from widely used method 
described in \cite{cow81}. In our method, the radial orbitals (RO) are obtained 
by solving the quasirelativistic equations of the following form:
\begin{eqnarray}
\left\{ \frac{d^2}{dr^2}-\frac{l(l+1)}{r^2}- V(nl|r)- \epsilon_{nl} 
\right\} P(nl|r) - X(nl|r) &
\nonumber \\
+ \frac{\alpha^2}{4} \left( \epsilon_{nl}+V\left(nl|r\right) \right)^2 P(nl|r) 
+ \frac{\alpha^2}{4} \left( \epsilon_{nl}+V(nl|r) \right) X(nl|r) &
\nonumber \\
+ \frac{\alpha^2}{4} \left( 1- \frac{\alpha^2}{4} \left( \epsilon_{nl}+V(nl|r) 
\right) \right)^{-1}D(nl|r)P(nl|r) & = 0.
\label{eq-qrhf}
\end{eqnarray}
The first line of Eq.(\ref{eq-qrhf}) represents the traditional Hartree-Fock 
equations, where $X(nl|r)$ denotes the exchange part of the potential and
$V(nl|r)$ represents the direct part of the potential including the interaction
of an electron with nucleus $U(r)$ and with other electrons. We take into 
account the finite size of a nucleus within the nucleus potential $U(r)$ 
\cite{bog02}. This allow us to represent the radial orbitals in powers of a 
radial variable in the nucleus region. Next two terms with the multiplier 
$\left( \epsilon_{nl} +V \left(nl|r\right)\right)$ describe the relativistic 
correction of the mass-velocity dependence. The last term of Eq.(\ref{eq-qrhf})
represents the potential of the electron contact interaction with the nucleus.
In our approach we include the contact interaction with the nucleus not only
for the s- electrons but also some part of it for the p- electrons:
\begin{equation}
D(nl|r) = \left( \delta(l,0)+\frac{1}{3}\delta(l,1) \right)
\frac{dU(r)}{dr} \left( \frac{d}{dr} - \frac{1}{r} 
\left( \alpha^2 Z^2 \delta(l,1) \left( - \frac{37}{30}
- \frac{5}{9n} + \frac{2}{3n^2} \right) + 1 \right) \right) .
\end{equation}
A detailed discussion of the particular features of Eq.(\ref{eq-qrhf}) is given 
in \cite{bog06, bog07}, whereas their solution techniques are described in 
\cite{bog03, bog05a}.

Concluding the description of applied approximation, we want to articulate the
unique features of our quasirelativistic method which significantly differs from
widely used approach described in \cite{cow81}. The main differences arise from 
our adopted set of Hartree-Fock quasirelativistic equations (QRHF) featuring 
several distinctive properties, namely: 

\begin{enumerate}[1.]
\item 
No statistical potentials are used. There are only conventional self-consistent 
field direct and exchange potentials in QRHF.
\item
The finite size of the nucleus is taken into account to determine the potential 
while solving the quasirelativistic equations \cite{bog02, bog05a}.
\item
The mass-velocity term is divided into two parts - the direct potential and the 
exchange one.
\item
The contact interaction term contains only the nucleus potential derivative in 
the numerator. There are no two-electron potentials in the numerator.
\item
Only the direct part of the potential is included into the denominator of the 
contact interaction with a nucleus term.
\item
The contact interaction with a nucleus is taken into account not only for 
s-electrons, but also for p-electrons with some additional corrections made
\cite{bog07}.
\end{enumerate}

One must look in \cite{bog06,bog07} for the most complete description of the way
to determine QRHF equations employed in this work. The methods to 
calculate the energy level spectra were discussed extensively in \cite{bog08}. 

For the energy level spectra calculation, we include all two-electron 
interactions in the same way as it is done in conventional Breit-Pauli 
approximation. This similarity makes it possible to apply widely
used code {\sc mchf breit} \cite{ah91} for the angular integration of the 
Breit-Pauli Hamiltonian matrix elements. We employ the computer program 
{\sc mchf mltpol} \cite{cff91} to determine the matrix elements of transition 
operators along with the code {\sc mchf lsjtr} \cite{cff91a} which has been 
adopted for use with the quasirelativistic radial orbitals.

\subsection{Correlation effects}{\label{tro}}

We use configuration interaction (CI) approximation to include electron 
correlation effects. In present work, depending on the type of one-electron 
state, two sorts of radial orbitals are exploited in the CI expansion of 
wavefunctions. For the configurations under consideration (adjusted 
configurations), we employ one-electron radial orbitals $P_{\mathrm{QR}}(nl|r)$
determined from solutions of Eq.(\ref{eq-qrhf}). This type of RO was applied for
all electrons with $n \leq 4$ and $l \leq 3$. We adopt the transformed radial 
orbitals (TRO) to describe virtually excited electrons included in CI expansion.

The method of transformed radial orbitals to account for the configuration 
interaction effects was developed in \cite{bog99} and successfully applied for 
the non-relativistic radial orbitals (see, e.g. \cite{bog05d, kar03, bog04a, 
bog09}). The comparison of TRO properties with those of the solutions of the 
multiconfigurational Hartree-Fock-Jucys equations \cite{juc52} demonstrates 
their close similarity \cite{bog04, bog05b}. Such a way to include correlation 
effects within our quasirelativistic approach was successfully applied for the 
studies of the tungsten spectral parameters in \cite{ran09}.

In current work the TRO were employed to describe the admixed configurations, 
having virtually excited electrons with the principal quantum number 
$4 < n \leq 7$ and with all allowed values of orbital quantum number $l$.
These orbitals are determined from the RO of the configurations under 
consideration by ensuring the effective inclusion of the correlation effects. 
For the present calculations, we employ the form of TRO with two free 
variational parameters $k$ and $B$:
\begin{equation}
P_{\mathrm{TRO}}(nl|r) = 
N \left( r^{l-l_0+k} \exp(-Br)P(n_0l_0|r) -
\sum_{n^{\prime} < n} P(n^{\prime}l|r) \int_{0}^{\infty} P(n^{\prime}l|r)
r^{l-l_0+k} \exp(-Br) P_{\mathrm{QR}}(n_0l_0|r)dr \right).
\end{equation}
Here the multiplier $N$ ensures the normalization of determined TRO, 
the first term in parenthesis performs the necessary transformation, and the
second term provides the orthogonality of all employed radial orbitals.  
The parameters $k$ and $B$ are varied in order to ensure the maximum of averaged
energy correction $\Delta E$ to the energy of the adjusted configuration in 
the second order of perturbation theory:
\begin{equation}
\Delta E_{\mathrm{PT}}(K_0,K^{\prime}) = 
\frac{\sum_{TLST^{\prime}}(2L+1)(2S+1) 
\langle K_0TLS \Vert H \Vert K^{\prime}T^{\prime}LS \rangle ^2}
{g(K_0) \left( {\bar E}(K^{\prime}) - {\bar E}(K_0) \right)}.
\end{equation}
We apply the same set of radial orbitals to describe both the even and odd 
configurations. This approach enables us to avoid any problems caused by the 
non-orthogonality of RO in calculation of the electron transition parameters.

As a consequence of adopted TRO, the basis of radial orbitals and the number of
possible admixed configurations increases rapidly.  Therefore one needs to 
perform a selection of the admixed configurations by including only those which
have the greatest influence on the configurations being adjusted. As a selection
criteria we apply the mean weight of the admixed configuration within the CI
expansion of the adjusted configuration wavefunction determined in the second
order of perturbation theory:
\begin{equation}
W_{\mathrm{PT}}(K_0,K^{\prime}) = 
\frac{\sum_{TLST^{\prime}}(2L+1)(2S+1) 
\langle K_0TLS \Vert H \Vert K^{\prime}T^{\prime}LS \rangle ^2}
{g(K_0) \left( {\bar E}(K^{\prime}) - {\bar E}(K_0) \right)^2}.
\label{eq-w}
\end{equation}

We include only those configurations which have 
$W_{\mathrm{PT}}(K_0,K^{\prime})$ larger than the specified small parameter $S$ 
in our calculation of energy levels and transition parameters. The methods and 
computer codes to calculate the mean weight from Eq.~(\ref{eq-w}) have been 
described in \cite{bog01, bog05c}. Such a way to select the interacting 
configurations in the non-relativistic approximation was sucessfully exploited 
(see, e.g., earlier mentioned \cite{bog05d, kar03, bog04a, bog09} and many 
others).

The configuration state functions (CSFs) of the selected configurations are used
to form the energy operator matrices. By diagonalizing these Hamiltonian 
matrices, we can determine the energy eigenvalues and eigenfunctions for the 
investigated configurations and exploit them for calculation of transition 
parameters. More details of employed approach one can find in \cite{bog08} and 
references therein, whereas some examples of application of our method are 
presented in \cite{ran09, kar10}.

In present work we have selected only those admixed configurations which 
have the mean weights in the wave functions of the configurations under 
consideration larger than $10^{-6}$. This procedure has allowed us to reduce the 
number of admixed configurations in the CI expansion of the wavefunctions to 102 
for even configurations and to 217 for odd configurations. So here we use an 
approximation denoted as {\bf F} in \cite{bog11}. It was demonstrated in 
\cite{bog11} that any further expansion of the admixed configurations basis does 
not lead to significant improvement in the calculated energy level spectra 
accuracy.

We have determined energy eigenvalues and eigenfunctions by diagonalizing 
Hamiltonian matrices obtained in the $LSJ$-coupling scheme in Breit-Pauli 
approximation. Using these results, we have calculated the transition 
parameters. We have computed data for the electric dipole (E1), electric 
quadrupole (E2) and magnetic dipole (M1) operators for the transitions both 
between the configurations under consideration and within these configurations.  
The emission transition rates of all above mentioned types were used to 
determine the radiative lifetimes of excited states.

\section{Results and discussion}{\label{result}}

We present the energy level spectra with the percentage contribution of 
eigenfunctions together with Lande factors $g$ and the radiative lifetimes
$\tau$ for the investigated configurations in Table~\ref{tab1}. It was 
demonstrated in \cite{bog11} that deviations of our theoretical level energy 
values from the experimental spectra data do not exceed $1\%$ in most cases. 
Consequently, we can conclude that these discrepancies in transition energies 
do not affect significantly the accuracy of calculated values of the transition 
rates or oscillator strengths. In Table~\ref{taben} we compare our calculated 
energy levels with data obtained from the completely relativistic calculations 
RELAC \cite{fou98} and with available experimental data \cite{utt02} presented
in compilation \cite{kra09}. This evaluation favorably demonstrates a better 
agreement of our data with the experimental level energies, especially for 
highly-excited states, comparing to the data from relativistic calculations 
RELAC. This is caused mainly by considerably improved inclusion of correlation 
effects.

Such a good agreement with experimental data leads to conclusion that our 
quasirelativistic approach adequately accounts for the relativistic effects 
even for highly-ionized heavy atoms. It is worth to mention a very good 
agreement of theoretical fine-structure level splitting for the ground term 
4p$^6$4d $^2$D pointing to close spin-orbit interaction values obtained in our 
work and in fully relativistic calculations \cite{fou98}.
 
As it was mentioned before, all our calculations were performed in $LS$-coupling
scheme. It is well established fact that the spin-orbit interaction induces
a very strong mixing of $LS$-terms for highly-charged ions. Therefore in  
Table~\ref{taben} we present only the level numbers (indices) $N$ taken from
Table~\ref{tab1} where one can find a corresponding CI wavefunction expansion.
It is easy to notice from the CSFs contributions presented in Table~\ref{tab1}
that the strong mixing occurs both for the $LS$-terms of the same configuration
and for terms of different configurations.  

A well-established case of a strong interaction between the 
$\mathrm{4p^54d^{N+1}}$ and $\mathrm{4p^64d^{N-1}4f}$ configurations (see 
\cite{jon10}) can be confirmed from the analysis of the wave function percentage
contributions. For example, a contribution from the $4p^64f$ configuration for 
the $J=2.5$ level $N=32$ in Table~\ref{tab1} is merely $49\%$. Moreover, the 
largest contribution for the level $N=28$ comes from the $\mathrm{4p^64f}$ 
configuration although the level itself clearly must be attributed to the 
configuration $\mathrm{4p^54d^2}$.

In Table~\ref{tab2} we present the transition wavelengths $\lambda$, the 
emission transition rates $A$ and the weighted oscillator strength values $gf$. 
We use only the level indices $N$ and the total angular momentum $J$ from 
Table~\ref{tab1} to describe both the initial and final levels along with the 
type of transition operator (E1, E2 or M1) applied to calculate the transition 
parameters. We present the transitions from the excited level to lower-lying 
levels in decreasing order of their rates. We limit the number of presented 
transitions by removing those having the transition rate values by 3 orders of 
magnitude lower comparing to the strongest transition from the particular energy 
level. This condition would remove completely the E2 transitions. Nonetheless, 
we have placed some most important E2 transitions in Table~\ref{tab2}. 

The magnetic dipole M1 transitions within configuration $\mathrm{4p^64d}$ were 
computed using a completely relativistic approximation in \cite{jon07}. 
As one can see from Table~\ref{tab2}, their value of transition rate 
$3.97 \times 10^4$ is very close to the value of $3.94 \times 10^4$ from our 
calculation.

From analysis of data presented in Table~\ref{tab2}, we can notice that the 
electric dipole transition rates are the most strong ones for all excited 
configuration levels when the total angular momentum $J \leq 3.5$. 
Furthermore, the strong E1 transitions define the values of radiative lifetimes 
$\tau$ for these levels presented in Table~\ref{tab1}. The only exemption from 
this rule demonstrates the level $N=14$ with total angular momentum $J=3.5$. 
Even if the E1 transitions from this level have the largest rates $A$, the 
values of two M1 transition rates are of the same order of magnitude as for the 
E1 transitions. This causes the significant decrease of the radiative lifetime 
value $\tau$ for this energy level.

The electric dipole transitions from the excited levels with $J \geq 4.5$ are 
forbidden because the largest value of total angular momentum of the ground 
state $J=2.5$. Consequently, the radiative lifetimes $\tau$ for these levels are
determined by the magnetic dipole transitions since the E2 transition rates are 
significantly smaller.

\ack
This work, partially supported by the European Communities under the contract 
of Association between EURATOM/LEI FU07-CT-2007-00063, was carried out within
the framework of the European Fusion Development Agreement.

\clearpage

\renewcommand{\baselinestretch}{1.0}
\begin{table}[hb!]
\caption{\label{taben}
Comparison of calculated energy levels (in 100 cm$^{-1}$) of W$^{37+}$ 
with available experimental data (EXP) from \cite{kra09} and 
relativistic calculations RELAC from \cite{fou98}. 
Level identifications from both \cite{kra09} and  \cite{fou98} are used where 
available, and level numbers $N$ are from Table~\ref{tab1}.
 	}
\begin{center}
\begin{tabular}{llrrrrr}
\hline\\
$N$ &
Term \cite{kra09} & 
Level Name \cite{fou98} & 
$J$ & 
EXP \cite{kra09} & 
QRHF & 
RELAC \cite{fou98}\\
\hline\noalign{\vskip4pt}
 1& 4p$^6$4d $^2$D	           & (4p)$^6$(4d$_-$)$^1$                             & 1.5 &     0 &     0 &     0 \\
 2& 4p$^6$4d $^2$D	           & (4p)$^6$(4d$_+$)$^1$                             & 2.5 &  1546 &  1538 &  1537 \\
 3& 4p$^5$4d$^2$ $^3$F (3/2,2)&                                                  & 0.5 & 12276 & 12311 &       \\
 4& 4p$^5$4d$^2$ $^3$F (3/2,2)&                                                  & 1.5 & 12276 & 12312 &       \\
 5&                           & (4p$_-$)$^2$(4p$_+$)$^3$(4d$_-$)$^2$&		            2.5 &  	    & 12591 & 12569 \\
 6&                           & (4p$_-$)$^2$(4p$_+$)$^3$(4d$_-$)$^2$&		            3.5 &	      & 12752 & 12720 \\
10&		                         & (4p$_-$)$^2$(4p$_+$)$^3$(4d$_-$)$^1$(4d$_+$)$^1$ & 4.5 &	      & 13922 & 13861 \\
13&		                         & (4p$_-$)$^2$(4p$_+$)$^3$(4d$_-$)$^1$(4d$_+$)$^1$ & 5.5 &	      & 14255 & 14201 \\
16&		                         & (4p$_-$)$^2$(4p$_+$)$^3$(4d$_-$)$^1$(4d$_+$)$^1$ & 3.5 &	      & 14631 & 14646 \\
17&		                         & (4p$_-$)$^2$(4p$_+$)$^3$(4d$_-$)$^1$(4d$_+$)$^1$ & 2.5 &	      & 14680 & 14691 \\
18&		                         & (4p$_-$)$^2$(4p$_+$)$^3$(4d$_-$)$^1$(4d$_+$)$^1$ & 1.5 &	      & 14756 & 14794 \\
19&		                         & (4p$_-$)$^2$(4p$_+$)$^3$(4d$_-$)$^1$(4d$_+$)$^1$ & 4.5 &	      & 14853 & 14861 \\
20& 4p$^5$4d$^2$ $^1$G (3/2,4)& (4p$_-$)$^2$(4p$_+$)$^3$(4d$_-$)$^1$(4d$_+$)$^1$ & 2.5 & 15085 & 15220 & 15220 \\
21&		                         & (4p$_-$)$^2$(4p$_+$)$^3$(4d$_+$)$^2$             & 5.5 &	      & 15500 & 15443 \\
22& 4p$^5$4d$^2$ $^3$F (3/2,3)& (4p$_-$)$^2$(4p$_+$)$^3$(4d$_-$)$^1$(4d$_+$)$^1$ & 1.5 & 15426 & 15618 & 15680 \\
24&	              	           & (4p$_-$)$^2$(4p$_+$)$^3$(4d$_+$)$^2$             & 3.5 &	      & 15671 & 15649 \\
25&		                         & (4p$_-$)$^2$(4p$_+$)$^3$(4d$_+$)$^2$             & 4.5 &	      & 15854 & 15851 \\
28&		                         & (4p$_-$)$^2$(4p$_+$)$^3$(4d$_+$)$^2$             & 3.5 &	      & 16561 & 16567 \\
30& 4p$^5$4d$^2$ $^3$F (3/2,4)& (4p$_-$)$^2$(4p$_+$)$^3$(4d$_+$)$^2$             & 2.5 & 17315 & 17535 & 17630 \\
32& 4p$^6$4f $^2$F	           & (4p)$^6$(4f$_-$)$^1$                             & 2.5 & 17581 & 17752 & 17840 \\
33& 4p$^6$4f $^2$F	           & (4p)$^6$(4f$_+$)$^1$                             & 3.5 & 17696 & 17889 & 17951 \\
34& 4p$^5$4d$^2$ $^3$F (1/2,2)& (4p$_-$)$^1$(4p$_+$)$^4$(4d$_-$)$^2$             & 2.5 & 20145 & 20319 & 20396 \\
44& 4p$^5$4d$^2$ $^1$D (1/2,2)&                                                  & 1.5 & 21843 & 22069 &       \\
46& 4p$^5$4d$^2$ $^1$G (1/2,4)& (4p$_-$)$^1$(4p$_+$)$^4$(4d$_-$)$^1$(4d$_+$)$^1$ & 3.5 & 23255 & 23274 & 23488 \\
47&		                         & (4p$_-$)$^1$(4p$_+$)$^4$(4d$_-$)$^1$(4d$_+$)$^1$ & 1.5 &	      & 23309 & 23525 \\
48&		                         & (4p$_-$)$^1$(4p$_+$)$^4$(4d$_-$)$^1$(4d$_+$)$^1$ & 2.5 &  	    & 23369 & 23347 \\
\hline\\
\end{tabular}
\end{center}
\end{table}

\clearpage				

\TableExplanation

\bigskip
\renewcommand{\arraystretch}{1.0}

\section*{Table 1.\label{tab1expl} 
Energy levels and radiative lifetimes of
$\mathbf{4p^64d, 4p^64f}$ and $\mathbf{4p^54d^2}$ configurations
of W$\mathbf{^{37+}}$.}

Throughout the table we present all the eigenfunction components which have
their percentage contributions higher than $10\%$

\begin{center}
\begin{tabular}{@{}p{1in}p{6in}@{}}
$N$          & The energy level number (index)\\
$E$          & The level energy in $10^2$ cm$^{-1}$\\
$g$          & The Lande factor $g$. The notation a(b) means $\mathrm{a \times 10^b}$\\
$\tau$       & The radiative lifetime in $10^{-9}$\,s \\
$J$          & The total angular momentum $J$ \\
Contribution & The percentage contribution of CSF in level eigenfunction\\
\end{tabular} 
\end{center}


\renewcommand{\arraystretch}{1.0}

\bigskip

\section*{Table 2.\label{tab2expl} 
Electron transitions among and within the $\mathbf{4p^64d}$,
$\mathbf{4p^64f}$ and $\mathbf{4p^54d^2}$ configurations of W$\mathbf{^{37+}}$.}

The transitions from the upper level to the lower ones are presented in the
descending order  of the transition rate values $A$. The number of presented 
transitions is limited to those having the transition rate values less than 
3 orders of magnitude smaller comparing to the strongest transition from the 
particular energy level.

\begin{center}
\begin{tabular}{@{}p{1in}p{6in}@{}}
$J_i$		& The total angular momentum $J$ of the initial level\\
$N_i$		& The initial energy level number $N$ \\
$J_f$		& The total angular momentum $J$ of the final level\\
$N_f$		& The final energy level number $N$ \\
Type            & The electron transition type \\
$\lambda$ (\AA) & The transition wavelength $\lambda$ in \AA \\
$A$ (s$^{-1})$  & The emission transition rate $A$ in s$^{-1}$. The notation aEb means $\mathrm{a \times 10^{b}}$\\
$gf$		& The weighted oscillator strength $gf$. The notation aEb means $\mathrm{a \times 10^{b}}$\\
\end{tabular}
\end{center}

\datatables 


\setlength{\LTleft}{0pt}
\setlength{\LTright}{0pt} 


\setlength{\tabcolsep}{0.7\tabcolsep}

\renewcommand{\arraystretch}{1.5}

\footnotesize 

\begin{longtable}{@{\extracolsep\fill}rrllllllll@{}}
\caption{\label{tab1}
Energy levels and radiative lifetimes of $\mathrm{4p^64d}$, $\mathrm{4p^64f}$ 
and  $\mathrm{4p^54d^2}$ configurations of W$^{37+}$. 
See page \pageref{tab1expl} for Explanation of Table.}
$N$ & 
\multicolumn{1}{c}{$E$} &
\multicolumn{1}{c}{$g$} & 
\multicolumn{1}{c}{$\tau$}& 
\multicolumn{1}{c}{$J$}& 
\multicolumn{5}{c}{Contribution}\\
\hline
\endfirsthead\\
\caption[]{(continued)}
$N$ & 
\multicolumn{1}{c}{$E$} &
\multicolumn{1}{c}{$g$} & 
\multicolumn{1}{c}{$\tau$}& 
\multicolumn{1}{c}{$J$}& 
\multicolumn{5}{c}{Contribution}\\
\hline \\
\endhead\\
1  &	    0 & 0.80 &  	         &1.5 & 98 4d                      $^2$D &        	                          &  		                               &			                                &		    		                           \\	      
2  &  1538 & 1.20 & 2.55($+$4) &2.5 & 98 4d                      $^2$D &        	                          &  		                               &			                                &	            	                     \\
3  & 12311 & 1.00 & 8.44($-$1) &1.5 & 42 4p$^5$4d$^2$($^3$F$_2$) $^4$D & +23 4p$^5$4d$^2$($^3$F$_2$) $^4$F & +10 4p$^5$4d$^2$($^1$D$_2$) $^2$D &			                                &			                                \\
4  & 12312 & 0.12 & 2.44($-$1) &0.5 & 84 4p$^5$4d$^2$($^3$F$_2$) $^4$D & +11 4p$^5$4d$^2$($^1$D$_2$) $^2$P &			                                &			                                &		    		                           \\
5  & 12591 & 1.03 & 3.83($-$1) &2.5 & 36 4p$^5$4d$^2$($^3$F$_2$) $^4$F & +18 4p$^5$4d$^2$($^1$D$_2$) $^2$D & +12 4p$^5$4d$^2$($^3$F$_2$) $^4$G & +10 4p$^5$4d$^2$($^3$F$^2$) $^4$D &		    		                           \\
6  & 12752 & 1.03 & 1.75($+$1) &3.5 & 32 4p$^5$4d$^2$($^3$F$_2$) $^4$G & +31 4p$^5$4d$^2$($^3$F$_2$) $^2$G & +23 4p$^5$4d$^2$($^1$D$_2$) $^2$F &  +7 4p$^5$4d$^2$($^3$F$_2$) $^4$F &		    		                           \\
7  & 13284 & 1.36 & 5.37($+$0) &1.5 & 28 4p$^5$4d$^2$($^3$F$_2$) $^4$P & +21 4p$^5$4d$^2$($^1$S$_0$) $^2$P & +15 4p$^5$4d$^2$($^3$P$_2$) $^4$D &  		                               & 		    		                          \\
8  & 13636 & 1.30 & 1.95($+$1) &2.5 & 32 4p$^5$4d$^2$($^3$F$_2$) $^4$D & +22 4p$^5$4d$^2$($^3$P$_2$) $^4$P & +15 4p$^5$4d$^2$($^3$P$_2$) $^4$D &  		                               & 		    		                          \\
9  & 13915 & 1.29 & 2.88($-$1) &3.5 & 35 4p$^5$4d$^2$($^3$F$_2$) $^4$F & +27 4p$^5$4d$^2$($^3$P$_2$) $^4$D & +16 4p$^5$4d$^2$($^3$F$_2$) $^4$D & +10 4p$^5$4d$^2$($^1$D$_2$) $^2$F &		    		                           \\
10 & 13922 & 1.18 & 3.04($+$4) &4.5 & 70 4p$^5$4d$^2$($^3$F$_2$) $^4$G & +15 4p$^5$4d$^2$($^3$F$_2$) $^2$G & +13 4p$^5$4d$^2$($^3$F$_2$) $^4$F &  		                               & 		    		                          \\
11 & 14052 & 1.09 & 1.29($-$1) &1.5 & 40 4p$^5$4d$^2$($^3$F$_2$) $^4$D & +14 4p$^5$4d$^2$($^3$F$_2$) $^2$D & +14 4p$^5$4d$^2$($^1$D$_2$) $^2$D & +11 4p$^5$4d$^2$($^1$D$_2$) $^2$P &		    		                           \\
12 & 14104 & 1.23 & 4.29($-$1) &2.5 & 40 4p$^5$4d$^2$($^3$P$_2$) $^4$P & +17 4p$^5$4d$^2$($^3$F$_2$) $^2$F & +15 4p$^5$4d$^2$($^1$G$_2$) $^2$F &  		                               & 		    		                          \\
13 & 14255 & 1.17 & 2.68($+$6) &5.5 & 57 4p$^5$4d$^2$($^1$G$_2$) $^2$H & +42 4p$^5$4d$^2$($^3$F$_2$) $^4$G &		   		                            &  		                               & 		    		                          \\
14 & 14290 & 1.06 & 1.15($+$4) &3.5 & 29 4p$^5$4d$^2$($^3$F$_2$) $^2$G & +15 4p$^5$4d$^2$($^3$F$_2$) $^2$F & +15 4p$^5$4d$^2$($^3$F$_2$) $^4$F & +15 4p$^5$4d$^2$($^1$G$_2$) $^2$G & +14 4p$^5$4d$^2$($^1$G$_2$) $^2$F \\
15 & 14408 & 1.66 & 4.37($+$2) &0.5 & 33 4p$^5$4d$^2$($^3$P$_2$) $^4$P & +29 4p$^5$4d$^2$($^3$P$_2$) $^2$S & +19 4p$^5$4d$^2$($^1$D$_2$) $^2$P &  		                               & 		                                \\
16 & 14631 & 1.17 & 3.20($+$1) &3.5 & 20 4p$^5$4d$^2$($^3$P$_2$) $^4$D & +13 4p$^5$4d$^2$($^1$D$_2$) $^2$F & +13 4p$^5$4d$^2$($^1$G$_2$) $^2$G & +12 4p$^5$4d$^2$($^1$G$_2$) $^2$F & +11 4p$^5$4d$^2$($^3$F$_2$) $^4$D \\
17 & 14680 & 1.22 & 1.60($-$1) &2.5 & 36 4p$^5$4d$^2$($^3$P$_2$) $^4$D & +25 4p$^5$4d$^2$($^3$P$_2$) $^2$D & +15 4p$^5$4d$^2$($^1$D$_2$) $^2$D &  		                               & 		    		                          \\
18 & 14756 & 1.39 & 2.77($+$0) &1.5 & 30 4p$^5$4d$^2$($^3$P$_2$) $^4$S & +25 4p$^5$4d$^2$($^3$P$_2$) $^2$D & +17 4p$^5$4d$^2$($^3$P$_2$) $^4$P &  		                               & 		    		                          \\
19 & 14853 & 1.13 & 1.47($+$5) &4.5 & 38 4p$^5$4d$^2$($^1$G$_2$) $^2$G & +24 4p$^5$4d$^2$($^3$F$_2$) $^4$F & +18 4p$^5$4d$^2$($^3$F$_2$) $^2$G & +17 4p$^5$4d$^2$($^1$G$_2$) $^2$H &		    		                           \\
20 & 15220 & 0.88 & 6.50($-$3) &2.5 & 41 4p$^5$4d$^2$($^1$G$_2$) $^2$F & +28 4f                      $^2$F & +13 4p$^5$4d$^2$($^3$F$_2$) $^2$F &  		                               & 		    		                          \\
21 & 15500 & 1.20 & 1.70($+$4) &5.5 & 57 4p$^5$4d$^2$($^3$F$_2$) $^4$G & +42 4p$^5$4d$^2$($^1$G$_2$) $^2$H &		        	                        &  		                               & 		    		                          \\
22 & 15618 & 0.85 & 2.16($-$3) &1.5 & 32 4p$^5$4d$^2$($^3$F$_2$) $^4$F & +22 4p$^5$4d$^2$($^3$P$_2$) $^2$D & +12 4p$^5$4d$^2$($^3$F$_2$) $^2$D &  		                               & 		    		                          \\
23 & 15619 & 1.21 & 1.60($+$0) &2.5 & 24 4p$^5$4d$^2$($^1$D$_2$) $^2$D & +18 4p$^5$4d$^2$($^3$P$_2$) $^4$P & +15 4p$^5$4d$^2$($^3$F$_2$) $^2$D & +13 4p$^5$4d$^2$($^3$P$_2$) $^2$D &		    		                           \\
24 & 15671 & 1.22 & 5.32($-$2) &3.5 & 34 4p$^5$4d$^2$($^1$D$_2$) $^2$F & +29 4p$^5$4d$^2$($^3$P$_2$) $^4$D & +13 4f$^2$F                       &  		                               & 		    		                          \\
25 & 15854 & 1.17 & 1.20($+$4) &4.5 & 38 4p$^5$4d$^2$($^3$F$_2$) $^4$F & +30 4p$^5$4d$^2$($^1$G$_2$) $^2$G & +18 4p$^5$4d$^2$($^3$F$_2$) $^2$G & +13 4p$^5$4d$^2$($^1$G$_2$) $^2$H &		    		                           \\
26 & 15972 & 1.08 & 6.58($-$3) &0.5 & 65 4p$^5$4d$^2$($^1$D$_2$) $^2$P & +14 4p$^5$4d$^2$($^3$P$_2$) $^4$P & +12 4p$^5$4d$^2$($^3$P$_2$) $^2$S &  		                               & 		    		                          \\
27 & 16350 & 0.76 & 1.49($-$3) &0.5 & 52 4p$^5$4d$^2$($^3$P$_2$) $^2$P & +24 4p$^5$4d$^2$($^3$P$_2$) $^4$D & +18 4p$^5$4d$^2$($^3$P$_2$) $^2$S &  		                               & 		    		                          \\
28 & 16551 & 1.18 & 4.48($-$2) &3.5 & 17 4f $^2$F                      & +16 4p$^5$4d$^2$($^3$F$_2$) $^2$F & +15 4p$^5$4d$^2$($^1$D$_2$) $^2$F & +14 4p$^5$4d$^2$($^3$F$_2$) $^4$D & +12 4p$^5$4d$^2$($^3$P$_2$) $^4$D \\
29 & 16897 & 1.27 & 5.82($-$1) &1.5 & 50 4p$^5$4d$^2$($^1$S$_0$) $^2$P & +14 4p$^5$4d$^2$($^3$P$_2$) $^2$D &		        	                        &  		                               & 		                                \\
30 & 17535 & 1.09 & 1.02($-$3) &2.5 & 29 4p$^5$4d$^2$($^3$F$_2$) $^2$D & +15 4p$^5$4d$^2$($^3$F$_2$) $^4$F & +13 4f  $^2$F                     &  		                               & 		                                \\
31 & 17553 & 1.36 & 1.58($-$3) &1.5 & 20 4p$^5$4d$^2$($^1$S$_0$) $^2$P & +17 4p$^5$4d$^2$($^3$P$_2$) $^4$S & +16 4p$^5$4d$^2$($^3$P$_2$) $^2$P & +13 4p$^5$4d$^2$($^1$D$_2$) $^2$D & +12 4p$^5$4d$^2$($^3$P$_2$) $^4$D \\
32 & 17752 & 0.93 & 1.21($-$3) &2.5 & 49 4f $^2$F                      & +18 4p$^5$4d$^2$($^1$G$_2$) $^2$F & 		   		                           &  		                               & 		                                \\
33 & 17889 & 1.12 & 1.72($-$3) &3.5 & 63 4f $^2$F                      & +18 4p$^5$4d$^2$($^1$G$_2$) $^2$G & 		   		                           &  		                               & 		                                \\
34 & 20319 & 0.77 & 8.80($-$4) &2.5 & 39 4p$^5$4d$^2$($^3$F$_2$) $^4$G & +28 4p$^5$4d$^2$($^3$F$_2$) $^2$F & +15 4p$^5$4d$^2$($^1$D$_2$) $^2$F &  		                               & 		                                \\
35 & 21314 & 1.02 & 7.13($-$1) &3.5 & 46 4p$^5$4d$^2$($^3$F$_2$) $^4$G & +19 4p$^5$4d$^2$($^3$F$_2$) $^2$G & +14 4p$^5$4d$^2$($^3$F$_2$) $^4$F &  		                               & 		                                \\
36 & 21387 & 1.07 & 1.70($-$2) &2.5 & 18 4p$^5$4d$^2$($^3$F$_2$) $^4$F & +16 4p$^5$4d$^2$($^1$D$_2$) $^2$F & +14 4p$^5$4d$^2$($^3$F$_2$) $^4$G & +14 4p$^5$4d$^2$($^3$P$_2$) $^2$D & +12 4p$^5$4d$^2$($^3$F$_2$) $^2$D \\
37 & 21478 & 1.07 & 9.10($-$3) &1.5 & 37 4p$^5$4d$^2$($^3$P$_2$) $^4$D & +16 4p$^5$4d$^2$($^3$F$_2$) $^2$D &		        	                        &  	  			                           &		     		                          \\
38 & 21547 & 1.96 & 1.39($-$3) &0.5 & 41 4p$^5$4d$^2$($^3$P$_2$) $^4$P & +33 4p$^5$4d$^2$($^3$P$_2$) $^2$S & +13 4p$^5$4d$^2$($^3$P$_2$) $^2$P &  	  			                           &		     		                          \\
39 & 21555 & 0.43 & 9.08($-$4) &0.5 & 56 4p$^5$4d$^2$($^3$P$_2$) $^4$D & +21 4p$^5$4d$^2$($^3$P$_2$) $^2$P & +13 4p$^5$4d$^2$($^1$S$_0$) $^2$P &  	  			                           &		     		                          \\
40 & 21754 & 1.04 & 1.39($+$2) &4.5 & 45 4p$^5$4d$^2$($^1$G$_2$) $^2$H & +19 4p$^5$4d$^2$($^1$G$_2$) $^2$G & +16 4p$^5$4d$^2$($^3$F$_2$) $^2$G &  	  			                           &		     		                          \\
41 & 22069 & 0.88 & 3.03($-$4) &1.5 & 43 4p$^5$4d$^2$($^3$F$_2$) $^2$D & +17 4p$^5$4d$^2$($^1$G$_2$) $^2$D & +12 4p$^5$4d$^2$($^3$F$_2$) $^4$F &  	  			                           &		     		                          \\
42 & 22794 & 1.29 & 2.73($-$3) &3.5 & 45 4p$^5$4d$^2$($^3$F$_2$) $^4$D & +32 4p$^5$4d$^2$($^3$F$_2$) $^2$F & +18 4p$^5$4d$^2$($^3$F$_2$) $^4$F &  	  			                           &		     		                          \\
43 & 23116 & 1.15 & 3.88($-$4) &2.5 & 34 4p$^5$4d$^2$($^3$F$_2$) $^2$D & +12 4p$^5$4d$^2$($^3$P$_2$) $^2$D & +10 4p$^5$4d$^2$($^3$F$_2$) $^4$D &  	  			                           &		     		                          \\
44 & 23159 & 1.20 & 4.30($-$3) &1.5 & 33 4p$^5$4d$^2$($^1$D$_2$) $^2$P & +26 4p$^5$4d$^2$($^1$D$_2$) $^2$D & +12 4p$^5$4d$^2$($^3$P$_2$) $^4$D & +11 4p$^5$4d$^2$($^3$P$_2$) $^4$P &		     		                          \\
45 & 23221 & 1.11 & 1.30($+$2) &4.5 & 31 4p$^5$4d$^2$($^3$F$_2$) $^2$G & +23 4p$^5$4d$^2$($^1$G$_2$) $^2$H & +17 4p$^5$4d$^2$($^3$F$_2$) $^4$G & +15 4p$^5$4d$^2$($^3$F$_2$) $^4$F & +11 4p$^5$4d$^2$($^1$G$_2$) $^2$G \\
46 & 23274 & 1.06 & 5.17($-$4) &3.5 & 53 4p$^5$4d$^2$($^1$G$_2$) $^2$F & +29 4p$^5$4d$^2$($^1$G$_2$) $^2$G &		        	                        &  		                               &            		                     \\
47 & 23309 & 1.48 & 4.69($-$4) &1.5 & 52 4p$^5$4d$^2$($^3$P$_2$) $^2$P & +20 4p$^5$4d$^2$($^3$P$_2$) $^4$S & +11 4p$^5$4d$^2$($^3$P$_2$) $^4$P &  		                               & 		                                \\
48 & 23369 & 1.11 & 4.68($-$1) &2.5 & 35 4p$^5$4d$^2$($^1$D$_2$) $^2$F & +24 4p$^5$4d$^2$($^3$P$_2$) $^2$D & +17 4p$^5$4d$^2$($^3$P$_2$) $^4$D &  		                               & 		                                \\
49 & 24058 & 0.67 & 3.13($-$2) &0.5 & 75 4p$^5$4d$^2$($^1$S$_0$) $^2$P & +13 4p$^5$4d$^2$($^3$P$_2$) $^4$D &			                                &			                                &		                                 \\
\hline \\
\end{longtable}

\newpage

\renewcommand{\arraystretch}{1.5}

\begin{longtable}{@{\extracolsep\fill}lrlrlrll@{}}
\caption{\label{tab2}
Electron transitions among and within the 4p$^6$4d, 4p$^6$4f and 4p$^5$4d$^2$ 
configurations of W$^{37+}$. See page\ \pageref{tab2expl} for Explanation of 
Table.}
$J_i$ & $N_i$ & $J_f$ & $N_f$ &
\multicolumn{1}{c}{Type} &
$\lambda$ (~\AA) & 
\multicolumn{1}{c}{$A$\ (s$^{-1}$)} & 
\multicolumn{1}{c}{$gf$}\\
\hline
\endfirsthead\\
\caption[]{(continued)}
$J_i$ & $N_i$ & $J_f$ & $N_f$ &
\multicolumn{1}{c}{Type} &
$\lambda$ (~\AA) & 
\multicolumn{1}{c}{$A$\ (s$^{-1}$)} & 
\multicolumn{1}{c}{$gf$}\\
\hline\\
\endhead\\
2.5 & 2  & 1.5 & 1  & M1 & 650.2 & 3.93E$+$04 & 1.49E$-$05 \\
    &	   & 1.5 & 1  & E2 & 650.2 & 1.89E$+$01 & 7.19E$-$09 \\
1.5 & 3  & 1.5 & 1  & E1 & 81.23 & 1.15E$+$09 & 4.56E$-$03 \\
    &	   & 2.5 & 2  & E1 & 92.83 & 3.38E$+$07 & 1.75E$-$04 \\
0.5 & 4  & 1.5 & 1  & E1 & 81.22 & 4.09E$+$09 & 8.09E$-$03 \\
2.5 & 5  & 1.5 & 1  & E1 & 79.42 & 2.37E$+$09 & 1.34E$-$02 \\
    &	   & 2.5 & 2  & E1 & 90.47 & 2.47E$+$08 & 1.82E$-$03 \\
3.5 & 6  & 2.5 & 2  & E1 & 89.18 & 5.72E$+$07 & 5.45E$-$04 \\
1.5 & 7  & 2.5 & 2  & E1 & 85.14 & 1.78E$+$08 & 7.74E$-$04 \\
    &	   & 1.5 & 1  & E1 & 75.28 & 8.25E$+$06 & 2.80E$-$05 \\
2.5 & 8  & 1.5 & 1  & E1 & 73.34 & 5.13E$+$07 & 2.48E$-$04 \\
    &	   & 2.5 & 2  & E1 & 82.66 & 3.11E$+$04 & 1.91E$-$07 \\
3.5 & 9  & 2.5 & 2  & E1 & 80.80 & 3.47E$+$09 & 2.72E$-$02 \\
4.5 & 10 & 3.5 & 6  & M1 & 854.6 & 3.29E$+$04 & 3.60E$-$05 \\
    &	   & 3.5 & 6  & E2 & 854.6 & 2.13E$+$00 & 2.33E$-$09 \\
1.5 & 11 & 1.5 & 1  & E1 & 71.17 & 5.76E$+$09 & 1.75E$-$02 \\
    &	   & 2.5 & 2  & E1 & 79.91 & 1.97E$+$09 & 7.56E$-$03 \\
2.5 & 12 & 2.5 & 2  & E1 & 79.58 & 2.31E$+$09 & 1.31E$-$02 \\
    &	   & 1.5 & 1  & E1 & 70.90 & 2.44E$+$07 & 1.11E$-$04 \\
5.5 & 13 & 4.5 & 10 & M1 & 2999  & 3.68E$+$02 & 5.95E$-$06 \\
    &	   & 3.5 & 6  & E2 & 665.1 & 5.94E$+$00 & 4.73E$-$09 \\
3.5 & 14 & 2.5 & 2  & E1 & 78.42 & 4.94E$+$04 & 3.65E$-$07 \\
    &	   & 2.5 & 5  & M1 & 588.5 & 2.32E$+$04 & 9.64E$-$06 \\
    &	   & 3.5 & 6  & M1 & 650.0 & 1.42E$+$04 & 7.21E$-$06 \\
    &	   & 2.5 & 8  & M1 & 1528  & 2.66E$+$02 & 7.44E$-$07 \\
0.5 & 15 & 1.5 & 1  & E1 & 69.41 & 2.25E$+$06 & 3.25E$-$06 \\
3.5 & 16 & 2.5 & 2  & E1 & 76.38 & 3.12E$+$07 & 2.18E$-$04 \\
2.5 & 17 & 1.5 & 1  & E1 & 68.12 & 6.25E$+$09 & 2.61E$-$02 \\
    &	   & 2.5 & 2  & E1 & 76.09 & 2.40E$+$04 & 1.25E$-$07 \\
1.5 & 18 & 2.5 & 2  & E1 & 75.65 & 2.47E$+$08 & 8.49E$-$04 \\
    &	   & 1.5 & 1  & E1 & 67.77 & 1.13E$+$08 & 3.12E$-$04 \\
4.5 & 19 & 3.5 & 6  & M1 & 475.9 & 3.48E$+$03 & 1.18E$-$06 \\
    &	   & 3.5 & 9  & M1 & 1066  & 2.55E$+$03 & 4.34E$-$06 \\
    &	   & 4.5 & 10 & M1 & 1074  & 4.18E$+$02 & 7.23E$-$07 \\
    &	   & 3.5 & 14 & M1 & 1777  & 1.63E$+$02 & 7.72E$-$07 \\
    &	   & 5.5 & 13 & M1 & 1673  & 1.07E$+$02 & 4.49E$-$07 \\
    &	   & 3.5 & 16 & M1 & 4510  & 4.65E$+$01 & 1.42E$-$06 \\
2.5 & 20 & 1.5 & 1  & E1 & 65.70 & 1.54E$+$11 & 5.96E$-$01 \\
    &	   & 2.5 & 2  & E1 & 73.09 & 1.94E$+$08 & 9.32E$-$04 \\
5.5 & 21 & 4.5 & 10 & M1 & 633.7 & 4.30E$+$04 & 3.11E$-$05 \\
    &	   & 5.5 & 13 & M1 & 803.5 & 1.50E$+$04 & 1.74E$-$05 \\
    &	   & 4.5 & 19 & M1 & 1546  & 5.80E$+$02 & 2.49E$-$06 \\
1.5 & 22 & 1.5 & 1  & E1 & 64.03 & 4.61E$+$11 & 1.13E$+$00 \\
    &	   & 2.5 & 2  & E1 & 71.02 & 1.64E$+$09 & 4.97E$-$03 \\
2.5 & 23 & 2.5 & 2  & E1 & 71.02 & 4.59E$+$08 & 2.08E$-$03 \\
    &	   & 1.5 & 1  & E1 & 64.02 & 1.66E$+$08 & 6.13E$-$04 \\
3.5 & 24 & 2.5 & 2  & E1 & 70.76 & 1.88E$+$10 & 1.13E$-$01 \\
4.5 & 25 & 3.5 & 14 & M1 & 639.4 & 3.06E$+$04 & 1.87E$-$05 \\
    &	   & 3.5 & 9  & M1 & 515.6 & 2.63E$+$04 & 1.05E$-$05 \\
    &	   & 4.5 & 10 & M1 & 517.5 & 1.50E$+$04 & 6.01E$-$06 \\
    &	   & 4.5 & 19 & M1 & 998.8 & 6.50E$+$03 & 9.73E$-$06 \\
    &	   & 5.5 & 13 & M1 & 625.5 & 3.67E$+$03 & 2.15E$-$06 \\
    &	   & 3.5 & 6  & M1 & 322.3 & 7.60E$+$02 & 1.18E$-$07 \\
    &	   & 3.5 & 16 & M1 & 817.7 & 5.13E$+$02 & 5.14E$-$07 \\
0.5 & 26 & 1.5 & 1  & E1 & 62.61 & 1.52E$+$11 & 1.79E$-$01 \\
0.5 & 27 & 1.5 & 1  & E1 & 61.16 & 6.71E$+$11 & 7.53E$-$01 \\
3.5 & 28 & 2.5 & 2  & E1 & 66.61 & 2.23E$+$10 & 1.19E$-$01 \\
1.5 & 29 & 2.5 & 2  & E1 & 65.11 & 1.23E$+$09 & 3.13E$-$03 \\
    &	   & 1.5 & 1  & E1 & 59.18 & 4.87E$+$08 & 1.02E$-$03 \\
2.5 & 30 & 1.5 & 1  & E1 & 57.03 & 5.33E$+$11 & 1.56E$+$00 \\
    &	   & 2.5 & 2  & E1 & 62.51 & 4.46E$+$11 & 1.57E$+$00 \\
1.5 & 31 & 2.5 & 2  & E1 & 62.44 & 6.34E$+$11 & 1.48E$+$00 \\
    &	   & 1.5 & 1  & E1 & 56.97 & 1.14E$+$08 & 2.21E$-$04 \\
2.5 & 32 & 1.5 & 1  & E1 & 56.33 & 5.39E$+$11 & 1.54E$+$00 \\
    &	   & 2.5 & 2  & E1 & 61.67 & 2.89E$+$11 & 9.89E$-$01 \\
3.5 & 33 & 2.5 & 2  & E1 & 61.16 & 5.81E$+$11 & 2.61E$+$00 \\
2.5 & 34 & 1.5 & 1  & E1 & 49.22 & 1.14E$+$12 & 2.47E$+$00 \\
    &	   & 2.5 & 2  & E1 & 53.25 & 7.48E$+$08 & 1.91E$-$03 \\
3.5 & 35 & 2.5 & 2  & E1 & 50.57 & 1.40E$+$09 & 4.28E$-$03 \\
2.5 & 36 & 2.5 & 2  & E1 & 50.38 & 5.28E$+$10 & 1.20E$-$01 \\
    &	   & 1.5 & 1  & E1 & 46.76 & 6.11E$+$09 & 1.20E$-$02 \\
1.5 & 37 & 1.5 & 1  & E1 & 46.56 & 9.49E$+$10 & 1.23E$-$01 \\
    &	   & 2.5 & 2  & E1 & 50.15 & 1.50E$+$10 & 2.26E$-$02 \\
0.5 & 38 & 1.5 & 1  & E1 & 46.41 & 7.20E$+$11 & 4.65E$-$01 \\
0.5 & 39 & 1.5 & 1  & E1 & 46.39 & 1.10E$+$12 & 7.11E$-$01 \\
4.5 & 40 & 5.5 & 13 & M1 & 133.4 & 4.27E$+$06 & 1.14E$-$04 \\
    &	   & 4.5 & 19 & M1 & 144.9 & 1.81E$+$06 & 5.68E$-$05 \\
    &	   & 3.5 & 16 & M1 & 140.4 & 3.26E$+$05 & 9.63E$-$06 \\
    &	   & 3.5 & 14 & M1 & 134.0 & 3.09E$+$05 & 8.32E$-$06 \\
    &	   & 3.5 & 9  & M1 & 127.6 & 8.37E$+$04 & 2.04E$-$06 \\
    &	   & 4.5 & 10 & M1 & 127.7 & 8.22E$+$04 & 2.01E$-$06 \\
    &	   & 5.5 & 13 & E2 & 133.4 & 6.84E$+$04 & 1.82E$-$06 \\
    &	   & 4.5 & 19 & E2 & 144.9 & 6.03E$+$04 & 1.90E$-$06 \\
1.5 & 41 & 1.5 & 1  & E1 & 45.31 & 3.29E$+$12 & 4.06E$+$00 \\
    &	   & 2.5 & 2  & E1 & 48.71 & 1.15E$+$10 & 1.63E$-$02 \\
3.5 & 42 & 2.5 & 2  & E1 & 47.05 & 3.67E$+$11 & 9.74E$-$01 \\
2.5 & 43 & 2.5 & 2  & E1 & 46.34 & 2.58E$+$12 & 4.98E$+$00 \\
    &	   & 1.5 & 1  & E1 & 43.26 & 6.71E$+$08 & 1.13E$-$03 \\
1.5 & 44 & 2.5 & 2  & E1 & 46.25 & 2.17E$+$11 & 2.78E$-$01 \\
    &	   & 1.5 & 1  & E1 & 43.18 & 1.57E$+$10 & 1.76E$-$02 \\
4.5 & 45 & 5.5 & 21 & M1 & 129.5 & 4.47E$+$06 & 1.13E$-$04 \\
    &	   & 4.5 & 25 & M1 & 135.7 & 2.10E$+$06 & 5.81E$-$05 \\
    &	   & 3.5 & 28 & M1 & 149.9 & 3.53E$+$05 & 1.19E$-$05 \\
    &	   & 3.5 & 33 & M1 & 187.5 & 1.57E$+$05 & 8.29E$-$06 \\
    &	   & 3.5 & 24 & M1 & 132.5 & 1.34E$+$05 & 3.52E$-$06 \\
    &	   & 4.5 & 25 & E2 & 135.7 & 8.96E$+$04 & 2.48E$-$06 \\
    &	   & 5.5 & 21 & E2 & 129.5 & 8.02E$+$04 & 2.02E$-$06 \\
    &	   & 3.5 & 35 & M1 & 524.5 & 7.74E$+$04 & 3.19E$-$05 \\
3.5 & 46 & 2.5 & 2  & E1 & 46.01 & 1.94E$+$12 & 4.91E$+$00 \\
1.5 & 47 & 2.5 & 2  & E1 & 45.93 & 2.13E$+$12 & 2.70E$+$00 \\
    &	   & 1.5 & 1  & E1 & 42.90 & 2.80E$+$09 & 3.09E$-$03 \\
2.5 & 48 & 2.5 & 2  & E1 & 45.81 & 2.12E$+$09 & 4.00E$-$03 \\
    &	   & 1.5 & 1  & E1 & 42.79 & 9.56E$+$06 & 1.57E$-$05 \\
0.5 & 49 & 1.5 & 1  & E1 & 41.57 & 3.20E$+$10 & 1.66E$-$02 \\
\hline\\
\end{longtable}


\begin{thebibliography}{99}
\bibitem{rea09}
        J. Reader,
        Phys. Scripta
        T134 (2009) 014023.
\bibitem{ski09}
        C. Skinner,
	       Phys. Scripta
        T134 (2009) 014022.
\bibitem{sum11}
        H. P. Summers, M. G. O'Mullane, 
        Am. Inst. Phys. Conf. Proc.  
        1344 (2011) 179.
\bibitem{bie09}
        C. Biedermann, R. Radtke, R. Seidel, T. P{\" u}tterich,
	       Phys. Scripta
        T134 (2009) 014026.
\bibitem{hut10}
	       R. Hutton, Y. Zou, M. Anderssin, T. Brage, I. Martinson, 
	       J. Phys. B: At. Mol. Opt. Phys.
	       43 (2010) 144026.
\bibitem{cle10}
	       J. Clementson, P. Beiersdorfer, E. W. Magee, H. S. McLean, R. D. Wood,
	       J. Phys. B: At. Mol. Opt. Phys.
	       43 (2010) 144009.
\bibitem{utt02}
        S. B. Utter, P. Beiersdorfer, E. Trabert,
        Can. J. Phys.
        80 (2002) 1503.
\bibitem{kra09}
        A. E. Kramida, T. Shirai, 
       	At. Data Nucl. Data Tables 
        95 (2009) 305.
\bibitem{fou98}
        K.B. Fournier,
       	At. Data Nucl. Data Tables 
        68 (1998) 1.
\bibitem{bog06}
        P. Bogdanovich, O. Rancova,
       	Phys. Rev. A
        74 (2006) 052501.
\bibitem{bog07}
        P. Bogdanovich, O. Rancova,
	       Phys. Rev. A
        76 (2007) 012507.
\bibitem{bog08}
        P. Bogdanovich, O. Rancova,
	       Phys. Scripta
        78 (2008) 045301.
\bibitem{bog10}
        P. Bogdanovich, V. Jonauskas, R. Karpu{\v s}kien{\. e}, O. Rancova,
	       Nuclear Instr. Meth. A
        619 (2010) 15.
\bibitem{bog11}
        P. Bogdanovich, O. Rancova, A. {\v S}tikonas,
	       Phys. Scripta
        83 (2011) 065302.
\bibitem{cow81}
        R. D. Cowan, {\em The Theory of Atomic Structure and Spectra}\/ 
	       University of California Press, Los Angeles, 1981
\bibitem{bog02}
        P. Bogdanovich, O. Rancova,
       	Lithuanian J. Phys. 
       	42 (2002) 257.
\bibitem{bog03}
        P. Bogdanovich, O. Rancova,
       	Lithuanian J. Phys. 
	       43 (2003) 177.
	\bibitem{bog05a}
        P. Bogdanovich, V. Jonauskas, O. Rancova,
       	Nuclear Instr. Meth. B
        235 (2005) 145.
\bibitem{ah91}
        A. Hibbert, R. Glass, C. F. Fischer,
       	Comput. Phys. Commun.
        64 (1991) 445.
\bibitem{cff91}
        C. F. Fischer, M. R. Godefroid, A. Hibbert,
       	Comput. Phys. Commun.
        64 (1991) 486.
\bibitem{cff91a}
        C. F. Fischer, M. R. Godefroid,
       	Comput. Phys. Commun.
        64 (1991) 501.
\bibitem{bog99}
        P. Bogdanovich, R. Karpu{\v s}kien{\. e},
       	Lithuanian J. Phys. 
	       39 (1999) 193.
\bibitem{bog05d}
        P. Bogdanovich, R. Karpu{\v s}kien{\. e}, I. Martinson,
        Phys. Scripta     
        67 (2003) 44.
\bibitem{kar03}
        R. Karpu{\v s}kien{\. e}, P. Bogdanovich,
	       J. Phys. B: At. Mol. Opt. Phys.
	       36 (2003) 2145.
\bibitem{bog04a}
        P. Bogdanovich, R. Karpu{\v s}kien{\. e}, A. Udris,
	       J. Phys. B: At. Mol. Opt. Phys.
	       37 (2004) 2067.
\bibitem{bog09}
        P. Bogdanovich, R. Karpu{\v s}kien{\. e},
       	At. Data Nucl. Data Tables 
        95 (2009) 533.
\bibitem{juc52}
        A. Yutsis,
       	JETP
        23, 2 (1952) 129 (in Russian).
\bibitem{bog04}
        P. Bogdanovich,
       	Lithuanian J. Phys. 
	       44 (2004) 135.
\bibitem{bog05b}
        P. Bogdanovich,
       	Nuclear Instr. Meth. B
        235 (2005) 92.
\bibitem{ran09}
        O. Rancova, P. Bogdanovich, R. Karpu{\v s}kien{\. e},
       	J. Phys.: Conf. Series
        163 (2009) 012011.
\bibitem{bog01}
        P. Bogdanovich, R. Karpu{\v s}kien{\. e},
       	Comput. Phys. Commun.
        134 (2001) 321.
\bibitem{bog05c}
        P. Bogdanovich, R. Karpu{\v s}kien{\. e}, A. Momkauskait{\. e},
       	Comput. Phys. Commun.
        172 (2005) 13.
\bibitem{kar10}
        R. Karpu{\v s}kien{\. e}, O. Rancova, P. Bogdanovich,
	       J. Phys. B: At. Mol. Opt. Phys.
	       43 (2010) 085002.
\bibitem{jon10}
        V. Jonauskas, R. Kisielius, S. Ku{\v c}as, A. Kynien{\. e},
       	Phys. Rev. A
        81 (2010) 012506.
\bibitem{jon07}
        V. Jonauskas, S. Ku{\v c}as, R. Karazija,
	       J. Phys. B: At. Mol. Opt. Phys.
        40 (2007) 2179.
\end{thebibliography}
\end{document}